\documentclass[superscriptaddress,secnumarabic,amssymb,amsmath,nobibnotes,aps,prd,showkeys,showpacs,nofootinbib,preprint]{revtex4}

\usepackage{graphicx}
\usepackage{float}
\usepackage{bm}
\usepackage{amsmath}
\usepackage{amsfonts}
\usepackage{amssymb}
\usepackage{epstopdf}
\usepackage{natbib}%
\newcommand{\bee}{\begin{equation}}
\newcommand{\eee}{\end{equation}}
\newcommand{\eaa}{\end{eqnarray}}
\newcommand{\baa}{\begin{eqnarray}}
\def\ni{\noindent}

\usepackage{color}

\begin{document}

\title{\Large Barrow black hole corrected-entropy model and Tsallis nonextensivity}

\author{Everton M. C. Abreu}\email{evertonabreu@ufrrj.br}
\affiliation{Departamento de F\'{i}sica, Universidade Federal Rural do Rio de Janeiro, 23890-971, Serop\'edica, RJ, Brazil}
\affiliation{Departamento de F\'{i}sica, Universidade Federal de Juiz de Fora, 36036-330, Juiz de Fora, MG, Brazil}
\affiliation{Programa de P\'os-Gradua\c{c}\~ao Interdisciplinar em F\'isica Aplicada, Instituto de F\'{i}sica, Universidade Federal do Rio de Janeiro, 21941-972, Rio de Janeiro, RJ, Brazil}

\author{Jorge Ananias Neto}\email{jorge@fisica.ufjf.br}
\affiliation{Departamento de F\'{i}sica, Universidade Federal de Juiz de Fora, 36036-330, Juiz de Fora, MG, Brazil}


\begin{abstract}
\ni The quantum scenario concerning Hawking radiation, gives us a precious clue that a black hole has its temperature directly connected to its area 
gravity and that its entropy is proportional to the horizon area. These results have shown that there exist a deep association between thermodynamics and gravity. 
The recently introduced Barrow formulation of back holes entropy, influenced by the spacetime geometry, shows the quantum fluctuations effects through Barrow exponent, $\Delta$, where $\Delta=0$ represents the usual spacetime and its maximum value, $\Delta=1$, characterizes a fractal spacetime. The quantum fluctuations are responsible for such fractality. 
Loop quantum gravity approach provided the logarithmic corrections to the entropy.   This correction arises from quantum and thermal equilibrium fluctuations.     
In this paper we have analyzed the nonextensive thermodynamical effects of the quantum fluctuations upon the geometry of a Barrow black hole. 
We discussed the Tsallis' formulation of this logarithmically corrected Barrow entropy to construct the equipartition law.  Besides, we obtained a master equation that provides the equipartition law for any value of the Tsallis $q$-parameter and we analyzed several different scenarios.   After that, the heat capacity were calculated and the thermal stability analysis was carried out as a function of the main parameters, namely, one of the so-called pre-factors, $q$ and $\Delta$.
\end{abstract}
\date{\today}
\pacs{04.70.Dy, 98.80.-k}
\keywords{Tsallis thermostatistics, Barrow entropy, logarithmically corrected-entropy approach, equipartition theorem}

\maketitle

\section{introduction}

The construction of an expression that represents the Bekenstein-Hawking area law for black hole (BH) entropy from the concept of the quantum geometry formalism \cite{abck}, together with an earlier idea, from string theory \cite{sv}, considering some peculiar cases, have had an increasing  interest concerning the quantum scenarios of BH physics in current times.


On the other hand, at present times, we are living under the aftermath of the observations of type Ia supernovae that  gave birth to dark energy concept.  Hence, both dark parts of the Universe are, the well known dark matter and the other one, as we mentioned, the dark energy \cite{varios-10}. The dark matter is considered a matter without pressure.   Its underlying function is to explain both the galactic rotation curves  and the development of large-scale framework.  Considering the second one,  which is a negative pressure exotic energy, it is utilized  to portray the current cosmic-accelerated expansion, although it is not known its origin and nature so far. But in spite of that, there exist several candidates that carry the concepts about its behavior and composition \cite{de-review}.


Although Barrow BH entropy was constructed as a toy model, there is some theoretical evidences that support his ideas.   In \cite{saridakis-1}, the authors constructed Barrow holographic energy.   They used the standard holographic principle at a cosmological structure and Barrow entropy, instead of the well known Bekenstein-Hawking one.   The authors demonstrated precisely that Barrow holographic dark energy can depict the Universe thermal history, with the sequence of matter  and dark energy eras \cite{saridakis-1}.  A dark energy EoS parameter was obtained where the $\Delta$-exponent affects this EoS and several dark energy scenarios were obtained as functions of $\Delta$-value.   In another support of Barrow's ideas, in \cite{saridakis-2}, the authors analyzed the validity of the generalized second law of thermodynamics using the Barrow entropy.   The sum of the entropy inside the apparent horizon plus the entropy of the horizon itself is always a non-decreasing function of time.   Hence, also concerning Barrow's entropy, the generalized second law is confirmed.   As another theoretical evidence, in \cite{saridakis-3}, the authors discussed modified cosmological scenarios using Barrow entropy,   The Friedmann equations were obtained when $\Delta=0$.   The new terms obtained constitute an effective dark energy sector.   They lead us to intriguing phenomenological behavior and, for $\Delta=0$, we observe that the $\Lambda$CDM concordance model is recovered.   Experimentally, in \cite{saridakis-4} the authors used observational data from Supernovae (SNIa) Pantheon sample, together with direct measurements of the Hubble parameter from the cosmic chronometers sample to obtain constraints concerning the scenario of Barrow holographic dark energy.   We strongly believe that, although being a toy model, all these evidences at least motivates, consistently, that Barrow's new concept of entropy deserves some profound investigation.

We have organized the ideas involved in this paper in the following way, in section 2 we have introduced the Barrow BH entropy through some recent results obtained by us.  In section 3 we have analyzed the Barrow BH entropy corrected logarithmically from the point of view of Tsallis thermostatistics.  We obtained a precise master equation that can provide us with the equipartition law for any $q$-parameter under different thermal scenarios. Besides, we computed the temperature, the number of degrees of freedom and the heat capacity to discuss the thermal stableness as functions of $q$ and $\Delta$. In section 4 we wrote the conclusions and some final remarks.

\section{Barrow black hole entropy and the equipartition law: previous results}


Recently, Barrow \cite{barrow-2}   analyzed the scenario where quantum gravitational effects could cause some  intricate, fractal structure on the BH surface.  It changes its actual horizon area, which in turn leads us to a new BH entropy relation, namely,
\bee
\label{barrow}
S_B\,=\,\bigg(\frac{A}{A_o} \bigg)^{1+\frac \Delta2}\,\,, 
\eee

\ni where $A$ is the usual horizon area  and $A_o$ the Planck area.   
The quantum gravitational perturbation is represented by the new exponent $\Delta$.  There are some characteristic values for $\Delta$.  For example, when $\Delta=0$  we have the simplest horizon construction.  In this case we obtain the well known  Bekenstein-Hawking entropy.   On the other hand, when   $\Delta=1$ we have the so-called  maximal deformation.  

Throughout this paper we will use the natural constant system where $\hbar=c=k_B=1$. In the context of the usual BH area entropy law, $S=A/4G$.  We will also assume that the number $N$ of degrees of freedom (dof) of the horizon satisfy the standard equipartition law \cite{pad}
\baa
\label{equi}
M= \frac{1}{2} N T \,,
\eaa

\ni where $T$ is the temperature, $M$ is the BH mass and we assume here that there are no interaction among the degrees of freedom of a BH. 

Our main target will be the Schwarszchild BH entropy, which characterizes the horizon.   Following Barrow deformed entropy, given in Eq. \eqref{barrow} for BH's \cite{barrow} it is given by
\bee
\label{a1}
S_B\,=\,\Bigg(\frac{A}{4G}\Bigg)^{1+\frac \Delta2}\,\,,
\eee

\ni where $G$ is the gravitation constant, $4G$ is the Planck area and $A$ is the standard horizon area.   
In BH physics, the area $A$ of the horizon can be associated with the source mass $M$ through the relation
\bee
\label{a2}
A=16 \pi G^2 M^2 \, .
\eee



\ni In \cite{nosso,nosso-1}, we substituted the area in Eq. \eqref{a2} into Eq. \eqref{a1} and we had that
\bee
\label{a4}
S_B\,=\,\Bigg(\frac{16\pi G^2 M^2}{4G}\Bigg)^{1+\frac \Delta2} \qquad \Longrightarrow \qquad S_B \,=\,\Big(4\pi G\Big)^{1+\frac \Delta2} M^{2+\Delta}\,\,.
\eee

The temperature is given by 
\bee
\label{a5}
\frac 1T=\frac{\partial S(M)}{\partial M} \,\,,
\eee

\ni where the expressions representing both BH entropy and temperature have a kind of universality, since both the horizon area and surface gravity are both completely geometric quantities, determined by the space-time geometry \cite{cai}.   Using Eq. \eqref{a5}
\bee
\label{a6}
T \,=\,  \frac{1}{\Big(2+\Delta \Big) \Big(4\pi G\Big)^{1+\frac \Delta2} M^{1+\Delta}}\,\,.
\eee

\ni We will use that the number of dof, $N$, in the horizon can be obtained by \cite{ko} 
\bee
\label{a7}
N\,=\,4\,S\,\,,
\eee

\ni where $S$ is the specific entropy that describes the horizon.   Hence, using Eqs. \eqref{a5} and \eqref{a7} we have that
\bee
\label{a8}
N \,=\, 4\Big(4\pi G\Big)^{\frac \Delta2+1} M^{\Delta+2}\,\,,
\eee

\ni but, substituting this result into Eq. \eqref{a6}, and solving the resulting equation for $M$, we have that
\bee
\label{a9}
M\,=\,\frac 12 \Big( 1\,+\,\frac \Delta2 \Big) N\,T\,\,,
\eee

\ni which corresponds to the horizon energy in Barrow's entropic formulation. From the last equation, if we make $\Delta = 0$, we recover the Bekenstein-Hawking equipartition law.

At this point, it is important to clarify what we are doing here exactly.  It is easy to understand that, for each entropy formulation we have two main quantities, namely, the number of dof and the equipartition law.   This last one is the energy of the event horizon.  The standard entropy is the Bekenstein-Hawking one.  Hence, we know that, for $N=4S$ dof we have that $M=1/2\,NT$.  So, for other formulations we have to change one of these two quantities.  Let us explain better by taking Barrow entropy for example.   It can be shown \cite{padmanabhan-dof} that its number of dof is $N=2(2+\Delta)\,S_{Barrow}$.   However, if we calculate the respective temperature, the final result is that $M=1/2\,NT$, independent of $\Delta$, and we have the same final result for any dof expression relative to a certain entropy.  But, on the other hand, if we keep the same number of dof relative to Bekenstein-Hawking entropy, of course we will have a new expression for the equipartition law.  And it is exactly what we are doing here, we are following this second path.   Namely, we will find a new equipartition law for each entropy formulation.   In other words, considering the same number of dof for Bekenstein-Hawking entropy, we will have different event horizon energies.   In one of our previous works \cite{nosso-2} we have obtained different horizon energies relative to different entropy formulations.


\section{Barrow nonextensive corrected-entropy and thermodynamics analysis}

The Bekenstein-Hawking entropy has an underlying r\^ole in holographic dark energy (HDE) model, where $S_{BH} = A/(4G)$ and it is used at the horizon \cite{wald}, as we said before. As a matter of fact, $A \sim L^2$ is the horizon's area.   Since the HDE model is connected to the area law of entropy, we have that a small adjustment to the area law of entropy will
change the HDE model energy density. One correction concerning the area law of entropy is the logarithmic correction \cite{varios-12} given by
\bee
\label{logaritmo} 
S_{BH} = \frac{A}{4G} + \widetilde{\alpha} \ln \frac{A}{4G} + \widetilde{\beta}\,\,,
\eee

\ni where  $\widetilde{\alpha}$ and $\widetilde{\beta}$  are dimensionless constant pre-factors and their values are still being discussed and not
yet confirmed even within LQG \cite{sf}. Several formulations concerning BH entropy provided the logarithmic
correction yielding $\widetilde{\alpha}= -1/2$ or $-3/2$  as standard values for this coefficient \cite{varios-13}. However, there is no such agreement
about regarding the way one might fix the value of the logarithmic pre-factor $\widetilde{\alpha}$, e.g., with $\widetilde{\beta} = 0$, because it seems to be a strongly model dependent parameter \cite{varios-14}.
The correction terms have an important and underlying function in both the late-time acceleration and early-time inflation of the Universe \cite{cll}.  It is easy to see that, for $\widetilde{\alpha}=\widetilde{\beta}=0$ we have Bekenstein-Hawking entropy.

As we said before, $\Delta$ represents the quantum fluctuations and the fractal feature of spacetime, which are motivated by LQG \cite{barrow-2}.   At the same time, we know that the introduction of quantum effects, motivated by LQG, caused by both thermal equilibrium and quantum fluctuations, conduct us to the curvature correction in Einstein's action known as the logarithmic entropy-correction described above in Eq. \eqref{logaritmo}.   Hence, we see that it is completely adequate to carry out the logarithmic correction concerning Barrow's entropy since both are connected by LQG fractal and quantum features.  

\subsection{Tsallis thermostatistics}

In \cite{tc} the authors analyzed the nonextensive generalization of the Bekenstein-Hawking entropy.   Hence, we will explore from now on the nonextensive generalization of the logarithmic-correction of both Bekenstein-Hawking and Barrow entropy.   In this nonextensive way, the expression introducing the $q$-logarithm for Bekenstein-Hawking entropy is given by
\bee
\label{A}
S_q\,=\,\frac{A}{4G}\,+\,\widetilde{\alpha} \ln_q \frac{A}{4G} \,+\,\widetilde{\beta}\,\,\,,
\eee

\ni where \cite{tc,tsallis}
\bee
\label{A1}
\ln_q x\,=\,\frac{x^{1-q}-1}{1-q}\,\,\,,
\eee

\ni and $q$ is the Tsallis parameter, which measures the nonextensivity, or extensivity, level of any specific system.   Hence, 
substituting Eqs. \eqref{a2} and \eqref{A1} into Eq. \eqref{A} we have that
\bee
\label{B}
S_q\,=\,4\pi G M^2\,+\,\frac{\widetilde{\alpha}(4\pi G)^{1-q}}{1-q}\,M^{2-2q}\,-\, \frac{\widetilde{\alpha}}{1-q} \,+\,\widetilde{\beta}\,\,.
\eee

\ni Using Eq. \eqref{a5} to calculate the temperature with Eq. \eqref{B}, it can be written as
\bee
\label{C}
T\,=\,\frac{1}{8\pi GM + 2\widetilde{\alpha} (4\pi G)^{1-q}\,M^{1-2q}}
\eee

The number of dof is
\bee
\label{D}
N\,=\,4\,\bigg[ 4\pi GM^2\,+\,\frac{\widetilde{\alpha}(4\pi G)^{1-q}}{1-q}\,M^{2-2q}\,-\, \frac{\widetilde{\alpha}}{1-q} \,+\,\widetilde{\beta}\bigg]\,\,\,.
\eee


Now, let us analyze the Barrow nonextensive corrected-entropy, but we will be back to this Bekenstein-Hawking corrected-entropy expression in a moment.  Remember that Bekenstein-Hawking entropy is Barrow one with $\Delta=0$.   Using again Eq. \eqref{a2}, the Tsallis logarithmically corrected entropy is 
\baa
\label{E}
S_q\,&=&\, \Bigg(\frac{A}{4G}\Bigg)^{1+\frac \Delta2}\,+\,\widetilde{\alpha} \ln_q \Bigg(\frac{A}{4G}\Bigg)^{1+\frac \Delta2}\,+\,\widetilde{\beta} \nonumber \\
\mbox{} \nonumber \\
&=& \Big(4\pi G\Big)^{1+\frac \Delta2} M^{2+\Delta}\,+\,\widetilde{\alpha}_{{}_{2}} \ln_q M\,+\,\widetilde{\beta}_{{}_{2}}\,\,\,,
\eaa

\ni where $\widetilde{\alpha}_{{}_{2}}=2\widetilde{\alpha}_{{}_{1}}$, $\widetilde{\beta}_{{}_{2}}=\widetilde{\beta}+\widetilde{\alpha}_{{}_{1}} \ln_q (4\pi G)$, 
$\widetilde{\alpha}_{{}_{1}} = \widetilde{\alpha} \Big(1+ \frac \Delta2 \Big)$ and substituting Eq. \eqref{A1} into Eq. \eqref{E} we have that
\bee
\label{F}
S_q\,=\,\Big(4\pi G\Big)^{1+\frac \Delta2} M^{2+\Delta}\,+\,\widetilde{\alpha}_{{}_{2}} \frac{M^{1-q}-1}{1-q} \,+\,\widetilde{\beta}_{{}_{2}}\,\,\,,
\eee

\ni and from Eq. \eqref{a5}, the temperature is given by
\bee
\label{G}
T_q\,=\,\frac{1}{\Big(4\pi G\Big)^{1+\frac \Delta2} \Big(2+\Delta\Big) M^{{}^{1+\Delta}}\,+\,\widetilde{\alpha}_{{}_{2}} \,M^{{}^{-q}}}\,\,\,.
\eee

Using again Eq. \eqref{D} to compute the number of dof, we can rewrite this equation such that
\bee
\label{H}
\frac N4 + \widetilde{\beta}_{{}_{3}}\,=\,\Big(4\pi G\Big)^{1+\frac \Delta2}\,M^{2+\Delta}\,+\,\widetilde{\alpha}_{{}_{3}}\,M^{1-q}\,\,\,,
\eee

\ni where $\widetilde{\alpha}_{{}_{3}}=\widetilde{\alpha}_{{}_{2}}/(1-q)$ and $\widetilde{\beta}_{{}_{3}} = \widetilde{\alpha}_{{}_{3}}\,-\,\widetilde{\beta}_{{}_{2}}$.   Eq. \eqref{G} will be also conveniently rewritten such that
\bee
\label{I}
\frac{M}{T_q}\,=\,\Big(4\pi G\Big)^{1+\frac \Delta2}\,\Big(2+\Delta\Big)\,M^{2+\Delta}\,+\,\widetilde{\alpha}_{{}_{2}}\,M^{1-q}\,\,\,.
\eee

\ni After some algebra combining Eqs. \eqref{H} and \eqref{I} we have a kind of master differential equation,
\baa
\label{J}
&&\hspace{2cm}\frac{M^q}{(2+\Delta)T_q}\,-\, \Big(\frac N4 + \widetilde{\beta}_{{}_{3}}\Big)\,M^{q-1}\,+\,\widetilde{\alpha}_{{}_{3}}\,-\,\widetilde{\alpha}_{{}_{2}}\,=\,0 \nonumber \\
\mbox{} \nonumber \\
\mbox{} \nonumber \\
&\!\!\!\!\!\!\Longrightarrow& \boxed{\frac{M^q}{(2+\Delta)T_q}\,-\,\bigg\{ \frac N4 + \bigg[ \frac{2}{1-q} - \ln_q (4\pi G)\bigg]\bigg(1+\frac \Delta2 \bigg) - \widetilde{\beta}\bigg\}M^{q-1}\,+\, \frac{2q\widetilde{\alpha}}{1-q}\bigg(1+\frac \Delta2\bigg)=0,} \nonumber \\
\eaa

\ni which can be understood as a master equation that can give us, as its solution, the form of the equipartition law, considering both Barrow entropy and Tsallis nonextensivity, for any $q$-parameter.   For example, if we substitute the limit $q \rightarrow 1$ and that $\widetilde{\alpha}=\widetilde{\beta}=\Delta = 0$, into Eq. \eqref{J} we have that $M=1/2\,NT$ as it should be, namely, the Bekenstein-Hawking entropy.  For $\widetilde{\alpha}=\widetilde{\beta}= 0$, i.e., neglecting the logarithmic correction we have that
\baa
\label{J-1}
&&\frac{M^q}{(2+\Delta)T}\,-\,\bigg\{ \frac N4 + \bigg[ \frac{2}{1-q} - \ln_q (4\pi G)\bigg]\bigg(1+\frac \Delta2 \bigg)\bigg\}M^{q-1}=0 \\
&\Longrightarrow& \boxed{M\,=\,\bigg\{ \frac N4 + \bigg[ \frac{2}{1-q}\,-\, \ln_q (4\pi G) \bigg]\bigg( 1+\frac \Delta2\bigg)\bigg\}\Big(1+\Delta\Big)\,T}
\eaa

For $\Delta=0$, we have the Bekenstein-Hawking nonextensive corrected-entropy, that can be written as
\bee
\label{K}
\boxed{\frac{M^q}{2T}\,-\,\bigg[\frac N4 + \widetilde{\alpha}\bigg(\frac{2}{1-q}\,-\,\ln_q 4\pi G \bigg)\,-\,\widetilde{\beta}\bigg]M^{q-1}\,+\,\frac{2\widetilde{\alpha}q}{1-q}\,=\,0 \,\,\,.}
\eee

\ni which is the nonextensivity expression for the logarithmically corrected Bekenstein-Hawking entropy.

From the master equation in Eq. \eqref{K} we see directly that 
for $q \in \mathbb{R}$ and $q\neq (1,2)$, we have to use numerical approaches to solve the master equation.   For $\widetilde{\alpha}=\widetilde{\beta}=0$, namely, no logarithmic correction, we have that, for any $q$ that
\bee
\label{K-1}
M^{q-1} \Big[\frac{M}{2T}\,-\,\frac N4 \Big]\,=\,0 \qquad \Longrightarrow \qquad M\,=\,\frac 12 NT\,\,\,,
\eee

\ni which is an interesting result.   We can say that the nonextensive version of Bekenstein-Hawking entropy is connected to the standard expression of the equipartition law independent of the value of $q$-parameter, when $\widetilde{\alpha}=\widetilde{\beta}=0$.   Although for the Bekenstein-Hawking corrected-entropy we have a strong dependence of $q$ to obtain the ``perturbed" equipartition law expression.

Back to Eq. \eqref{J} for $q=2$ we have that
\bee
\label{L}
M_{q=2}\,=\,\frac{(2+\Delta)T}{2}\,\Bigg[\frac N4+\widetilde{\beta}_{{}_{3}}\,\pm\,\sqrt{\bigg(\frac N4+\widetilde{\beta}_{{}_{3}}\bigg)^2
\,-\,4\Big(2+\Delta\Big)\Big(\widetilde{\alpha}_{{}_{3}}\,-\,\widetilde{\alpha}_{{}_{2}}\Big)T}\;\Bigg] \,\,\,,
\eee

\ni and for $\Delta=0$ we can write that
\bee
\label{L-1}
M_{q=2}^{\Delta=0}\,=\,T\,\Bigg[\frac N4\,+\,\widetilde{\beta}_{{}_{4}}\,\pm\,\sqrt{\bigg(\frac N4+\widetilde{\beta}_{{}_{4}}\bigg)^2\,+\,32\,\widetilde{\alpha}}\Bigg]\,\,\,,
\eee

\ni where $\widetilde{\beta}_{{}_{4}}=\widetilde{\beta}_{{}_{3,\Delta=0}}=-\,\widetilde{\alpha}\Big[1\,+\,\ln_{{}_{q=2}}\Big(4\pi G \Big) \bigg]$.  And, as expected for the positive root, $M=1/2 NT$ when $\widetilde{\alpha}=\widetilde{\beta}=\Delta=0$.
In Eq. \eqref{J}, for $\widetilde{\alpha}=\widetilde{\beta}= 0$, we have that, for any $q$-parameter, that 
\bee
\label{L-1}
M\,=\,\frac 12 \Big(1+\frac \Delta2 \Big)NT\,\,\,,
\eee

\ni which is Barrow equipartition theorem, obtained in \cite{nosso}.  Notice that Eq. \eqref{L-1} was obtained for any value of $q$.

The next step on our thermodynamics analysis is to compute the heat capacity,
\bee
\label{M}
C\,=\,-\,\frac{\Big[S'(M)\Big]^2}{S''(M)}\,\,,
\eee

\ni and to provide a thermodynamical analysis of the result, substituting Eq. \eqref{F} in Eq. \eqref{M} we have that
\bee
\label{N}
C_q\,=\,\frac{\Big[\Big(4\pi G\Big)^{1+\frac \Delta2} \Big(2+\Delta\Big)\,M^{1+\Delta}\,+\,\widetilde{\alpha}_{{}_{2}}\,M^{-q}\Big]^2}
{\Big(4\pi G\Big)^{1+\frac \Delta2} \Big(2+\Delta\Big)\Big(1+\Delta\Big) M^{{}^{\Delta}}\,-\,\widetilde{\alpha}_{{}_{2}} q M^{-1-q}}\,\,\,,
\eee

\ni and for thermodynamical stability
\bee
\label{O}
\widetilde{\alpha} > \frac{\Big(4\pi G\Big)^{1+\frac \Delta2}\Big(1+\Delta \Big) M^{1+q+\Delta}}{q}\,\,\,.
\eee

\ni and when $q=1$ in Eq. \eqref{O} we have the same result as in \cite{nosso}.



\section{Conclusions and last words}

Barrow's toy model for BH entropy comes from the fact that the BH area can be perturbed by the quantum gravitational effects.  
Namely, we could expect that quantum fluctuations from space-time can cause a modification of the topology of space-time at the Planck scale.  The result would be a foam-like framework called the space-time foam.   Hence, considering Barrow's formulation, we can measure its deviation from the Bekenstein-Hawking entropy through a new exponent $\Delta$, where $\Delta = 0$ means Bekenstein-Hawking entropy, and $\Delta=1$ means the most intricate case. We can see clearly that considering the standard Barrow entropy expression, when the fractal effect factor $\Delta$ is zeroed, i.e., when  it is withdrawn from the expression in Eq. \eqref{a9}, we obtain the standard equipartition theorem, which is a classical expression.   

In this work  
we analyzed the Tsallis nonextensive version of the logarithmically corrected Barrow entropy.   The objective is to obtain the respective modified equipartition law for the characteristic entropy.   The equipartition law corresponds to the horizon energy of barrow BH.

To sum up the results obtained here, we obtained in Eq. (22) a master equation with all the parameters of the model, i.e., $q$, $\Delta$, $\widetilde{\alpha}$ and $\widetilde{\beta}$.    This master equation is the source of the equipartition law expressions for any value of $q$.   After that, in Eq. (23), we computed the master equation without the logarithmically correction, namely, the nonextensive equation for Barrow entropy.   This $\widetilde{\alpha}=\widetilde{\beta}=0$ equation has the Barrow-Tsallis equipartition law as the solution and it is shown in Eq. (24).   In Eq. (25), we kept the logarithmic correction but now $\Delta=0$, which means that Eq. (25) is the master equation for the nonextensive logarithmic correction of Bekenstein-Hawking entropy.   Its solution, i.e., the energy of the Bekestein-Hawking BH horizon, is given in Eq. (26).  It is a function of the $q$-parameter, of course.

As an example of an analytic exact solution of Eq. (22), we presented the solution for $q=2$ in Eq. (27).  And its $\Delta=0$, Bekenstein-Hawking version in Eq. (28).  For any other value of $q$ different from $q=1$ and $q=2$, the complete equation needs numerical computation.  Of course the master equation allows the construction of any $q \in \mathbb{R}$ equipartition law, but again, numerical computation is mandatory.

Back to Eq. (22), the complete master equation, eliminating the logarithmic correction, i.e., $\widetilde{\alpha}=\widetilde{\beta}=0$, it makes the equation to be solved and hence, we obtained the Barrow equipartition law, in Eq. (29), which reproduced the result obtained in \cite{nosso}.  The curious fact is that Barrow equipartiion law is obtained regardless the value of $q$, as demonstrated in the algebraic work from Eq. (22) to obtain Eq. (29).

After all that,  to verify the thermodynamical stableness of the model, we have calculated the heat capacity of the system, which must be a positive quantity.   We have analyzed the positivity condition based upon the value of the $\widetilde{\alpha}$-parameter, since any derivative of the entropy make the constant $\widetilde{\beta}$-parameter disappear.   
We obtained the conditions on $\widetilde{\alpha}$, which confirms that the value of this pre-factor is dependent of the model.

\section*{Acknowledgments}

\ni The authors thank CNPq (Conselho Nacional de Desenvolvimento Cient\' ifico e Tecnol\'ogico), Brazilian scientific support federal agency, for partial financial support, Grants numbers  406894/2018-3 (E.M.C.A.) and 303140/2017-8 (J.A.N.).

\end{document}